\documentclass[a4paper,12pt]{article}
\usepackage[super,square]{natbib}
\usepackage{anysize}
\marginsize{2cm}{2cm}{1cm}{2cm}
\usepackage{fancyhdr}

\usepackage{soul}
\usepackage[dvipsnames]{xcolor}
\usepackage{hyperref}
\hypersetup{
    colorlinks=true,
    linkcolor=black,
    citecolor=CadetBlue,
    filecolor=CadetBlue,      
    urlcolor=CadetBlue,
}
\usepackage{multicol}
\usepackage{parskip}
\renewenvironment{abstract}
 {\par\noindent\textbf{\abstractname}\ \ignorespaces \\}
 {\par\noindent\medskip}

\usepackage{graphicx}

\usepackage{booktabs}
 
\begin{document}
\pagestyle{fancy}
\thispagestyle{empty}

\fancyhead[L]{}
\renewcommand*{\thefootnote}{\fnsymbol{footnote}}
\begin{center}
\Large{\textbf{Can AI Assist in Olympiad Coding?}}
\vspace{0.4cm}
\normalsize
\\ Samuel Ren \\
\vspace{0.1cm}
\textit{Henry Gunn High School}
\medskip
\normalsize
\end{center}
{\color{gray}\hrule}
\vspace{0.4cm}
\begin{abstract}
As artificial intelligence programs have become more powerful, their capacity for problem-solving continues to increase, approaching top-level competitors in many olympiads. Continued development of models and benchmarks is important but not the focus of this paper.  While further development of these models and benchmarks remains critical, the focus of this paper is different: we investigate how AI can {\textit{assist}} human competitors in high-level coding contests. In our proposed workflow, a human expert outlines an algorithm and subsequently relies on an AI agent for the implementation details. We examine whether such human–AI collaboration can streamline the problem-solving process and improve efficiency, highlighting the unique challenges and opportunities of integrating AI into competitive programming contexts.

\end{abstract}
{\color{gray}\hrule}
\medskip

\section{Introduction}
Over the last decade, the application of artificial intelligence (AI) to software engineering and programming education has grown exponentially. Advances in deep learning have led to AI models capable of generating code automatically, significantly reducing the manual effort required for software development. These systems, initially focused on simpler coding tasks such as writing boilerplate code or performing routine refactoring, have rapidly progressed to solving more challenging problems. Recent AI-powered code-generation platforms, such as OpenAI Codex~\cite{chen2021evaluatinglargelanguagemodels} and DeepMind’s AlphaCode~\cite{Li2022}, have demonstrated that large language models (LLMs) trained on massive datasets of source code can not only write correct solutions to many standard programming tasks, but also approach near-human or even superhuman performance on certain competitive programming problems.

Despite these achievements, it remains an open question which levels of coding competitions AI systems can reliably handle. Competitive programming tasks, particularly from high-level contests like the USA Computing Olympiad (USACO), often require not just algorithmic knowledge and puzzle solving but also creativity and efficient implementation. Recently, Wang \emph{et al.}~\cite{shi2024languagemodelssolveolympiad} introduced the USACO benchmark with 307 problems, accompanied by detailed tests, reference solutions, and official analyses. Their findings show that even powerful models like GPT-4 achieve only an 8.7\% pass@1 accuracy using zero-shot chain-of-thought prompting, and the best inference method tested reaches 20.2\% accuracy by integrating techniques such as self-reflection and retrieval. In addition, a small amount of human-driven guidance (in the form of hints) can significantly improve GPT-4's performance on problems previously unsolved by any model or method. These results underscore that state-of-the-art LLMs, while impressive, still face substantial difficulties on complex algorithmic tasks.

Beyond direct competition, AI has also been studied as a tool to enhance coding education and practice. Multiple research initiatives have investigated how AI-driven tutoring systems can improve learning outcomes by providing real-time feedback, hints, and personalized learning pathways. For aspiring programmers, such systems reduce the cognitive load associated with debugging and encourage active problem-solving. Moreover, widely available AI assistants like GitHub Copilot have been reported to boost developer productivity~\cite{chen2021evaluatinglargelanguagemodels}. While most existing studies focus on novice programmers, the benefits of AI assistance for more experienced coders---particularly those competing at a high level---are not yet thoroughly explored.

Therefore, in this work, we seek to address a key question: 

\textbf{Q: Can AI help experienced coding competition participants accelerate their problem solving?} 

Specifically, rather than having AI  solve entire challenges, our approach leverages a c\textbf{ollaborative paradigm} where a human competitor devises the algorithmic strategy but then consults an AI assistant to implement and refine the solution. By combining human insights on problem decomposition with AI’s speed and breadth of coding expertise, we hypothesize that this hybrid approach may improve problem-solving efficiency while still preserving the creative and analytical aspects of algorithm design.

To test this hypothesis, we have designed an experiment that compares the performance of experienced competitors on a set of tasks both \emph{with} and \emph{without} an AI assistant. By quantifying the metric on time to solution, we aim to offer insight into how AI assistance might affect high-level competitive programming outcomes. In addition, this study lays groundwork for future research on integrating AI more seamlessly into competitive coding workflows and helping participants---from novices to experts---boost their performance through advanced yet collaborative tools.

This work represents a first step toward understanding how experienced coding competition participants can benefit from AI-assisted solutions. There is much work to be done for a more comprehensive evaluation, from novices to experts, using questions of different difficulty levels, having a larger sample size, etc. 

Last, the work brings to the forefront several critical questions and potential directions for future research. In the short term, AI’s ability to generate complex code challenges the integrity of competitions like USACO, as participants may rely on AI tools to gain unfair advantages. To prevent misuse, we currently keep our AI-assisted workflow closed-source. Over the long run, coding competitions may need to  evolve in order to stay relevant, engaging, and educational,   as AI becomes an increasingly powerful  tool.

\section{Related Work}

Generative AI tools for coding have advanced significantly in recent years. In \cite{chen2021evaluatinglargelanguagemodels}, GitHub Copilot powered by Codex is evaluated and shown impressive improvement over GPT-3 and GPT-J.  
In \cite{Li2022}, the authors introduce AlphaCode for competition-Level coding. In simulated evaluations using the Codeforces platform, AlphaCode achieved on average a ranking of top 54.3\% generation. 

In \cite{becker2023programming}, challenges and opportunities in AI for coding in the context of education are discussed.  

Our work is inspired by \cite{shi2024languagemodelssolveolympiad} that introduced the USACO benchmark with 307 problems, and evaluated AI coding abilities on these problems. We build upon the prompts provided in the paper, as well as an updated version of their text corpus for Retrieval-Augmented Generation. The main difference is the question we asked in this work. While \cite{shi2024languagemodelssolveolympiad} aims to solve an entire problem, we focus on whether and how much can AI tools help human coders in solving the problem. 

\section{Methods}
We create a workflow to maximize efficiency in letting human and AI work together. 

\begin{figure}[h]
    \centering
    \includegraphics[width=\textwidth]{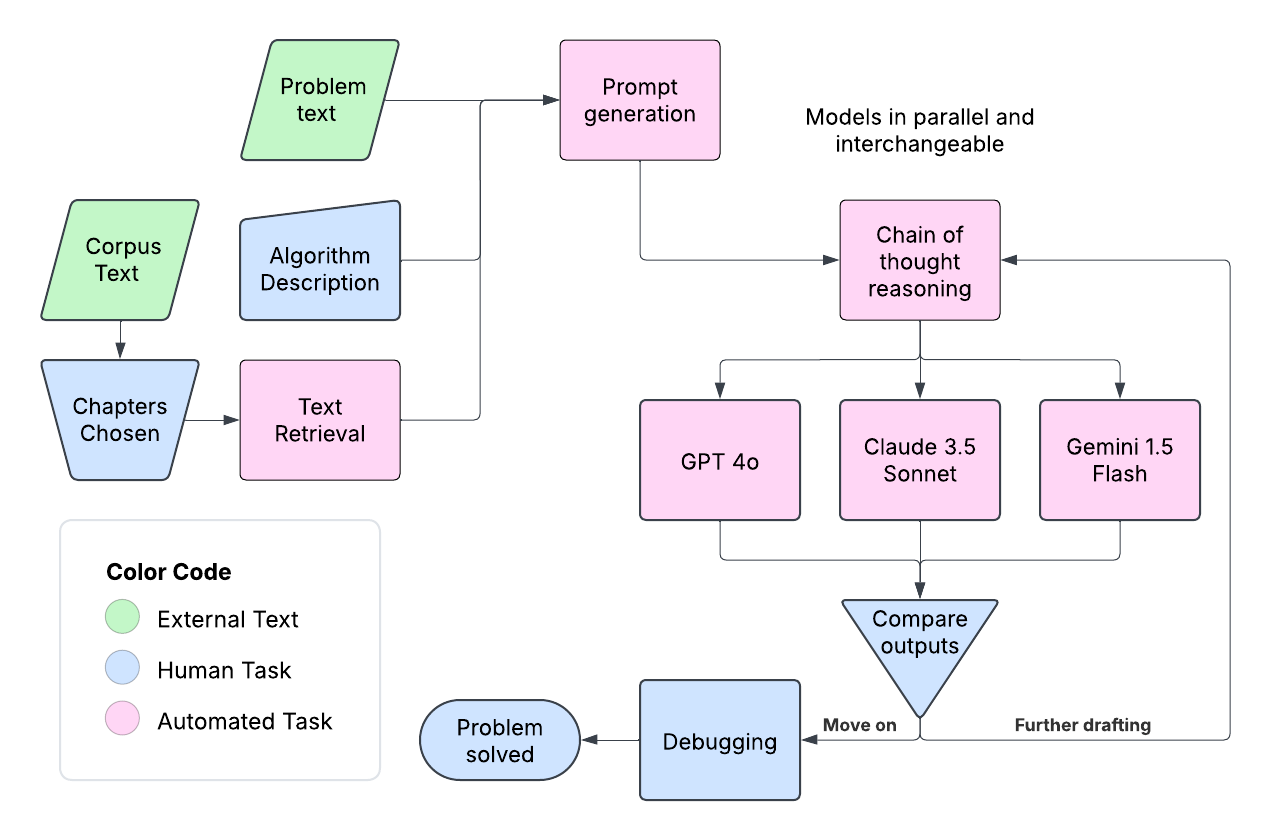} \caption{Workflow graph}
\end{figure}

The code generates a GUI for interacting with multiple models at once, including models from OpenAI, Anthropic, and Gemini. By default, it uses GPT 4o, Claude 3.5 Sonnet, and Gemini 1.5 Flash. Due to monetary constraints, OpenAI's O1 model is not available. A GUI is created with Tkinter to sort information and streamline usage. Models are instructed to code in C++, since olympiad programming almost always guarantees a C++ solution but not necessarily in other languages.

Since the models' providers have different APIs, a unified client normalizes the format within the code. Since models run at different speed, it utilizes the async capabilities of these APIs to update the GUI as information comes in. This client also stores message history to allow for chats between the user and the models. A new instance of the client is initialized for each model. It will also limit token usage to a constant set for each model. 

The upper half consists of human input and the bottom half is split into columns, each for a different model's output. 

To initiate the chats, the human provides three types of information:
1. The problem text. A button takes the text from the clipboard and stores it, allowing for quick input. The program requires the problem text before allowing any generation. 
2. The algorithm description. The user freely types into the main input text box for this section. The program does not require an algorithm to start.
3. The reference material. The human lists chapters of the CP-Algorithm website to feed into the model as reference material. The user inputs simplified file paths in a small text box. The program does not require reference material to start.

\begin{figure}[h]
    \centering
    \includegraphics[width=\textwidth, trim=0 500 0 0, clip]{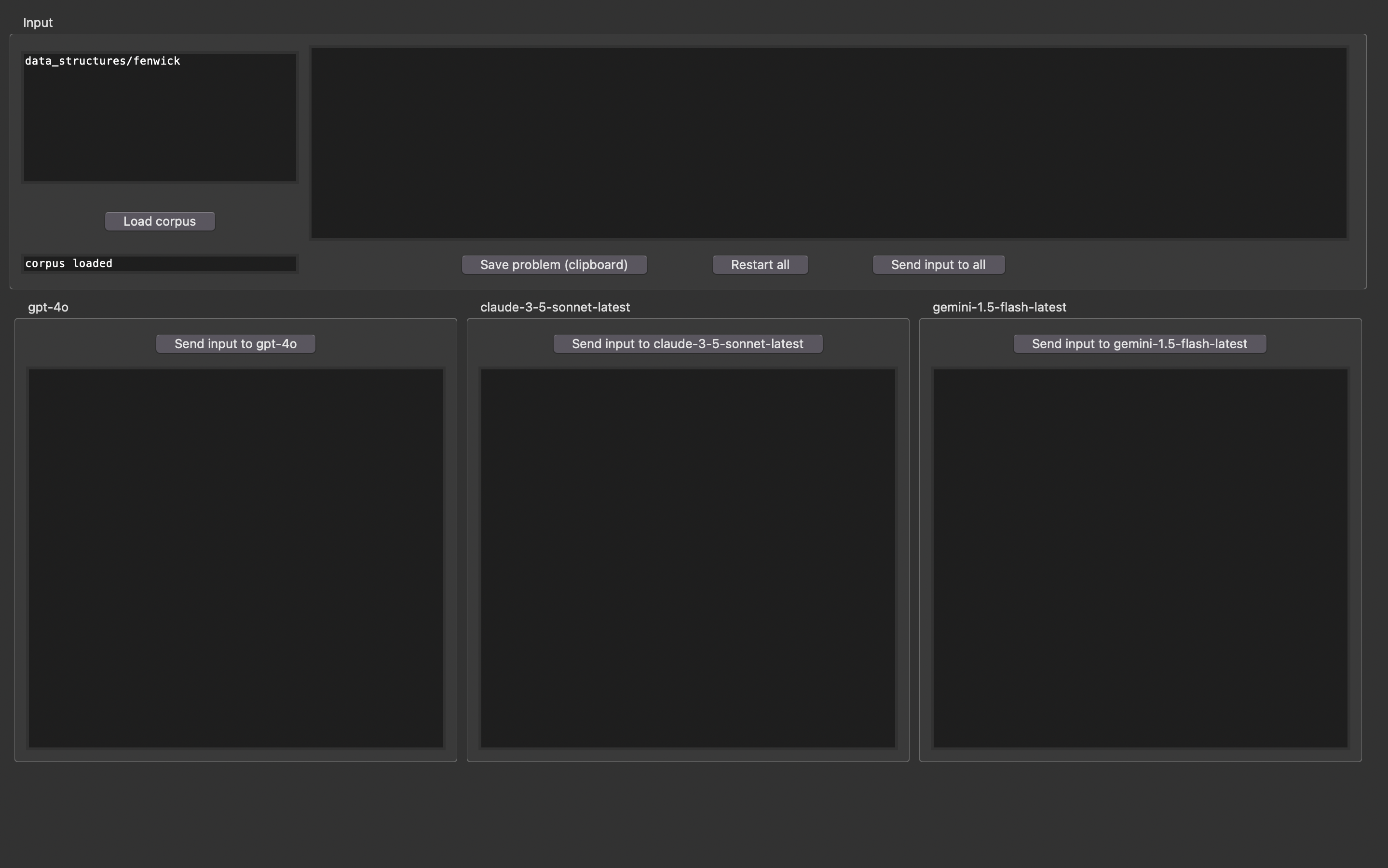} 
    \caption{Corpus loaded}
\end{figure}

\begin{figure}[h]
    \centering
    \includegraphics[width=\textwidth, trim=0 500 0 0, clip]{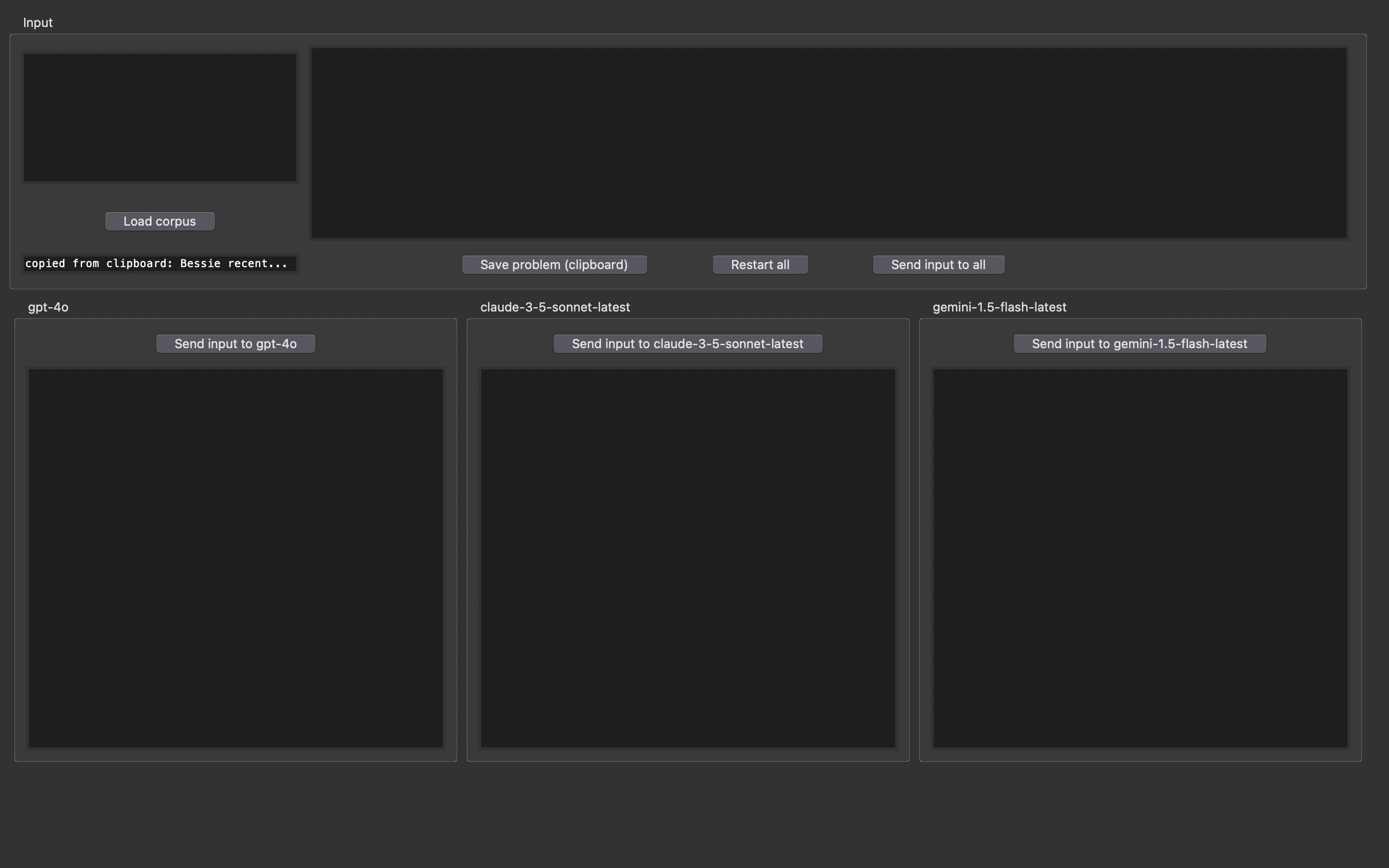} 
    \caption{Problem text loaded}
\end{figure}

The figures are trimmed for enlarged size. Upon inputting the problem text or corpus, a status message confirms success or reports an error. Once everything is loaded, the information is compiled into a prompt and sent to each model. 

A button sends the starting prompt to each of the models and updates the GUI as each API returns text. 

\newpage

Once the chats are all initiated, the main input text box is no longer used for the algorithm description, but becomes a place to write messages to the models. There is a button for each individual model and one to send to all of them at once. 

\begin{figure}[h]
    \centering
    \includegraphics[width=\textwidth, trim=0 100 0 250, clip]{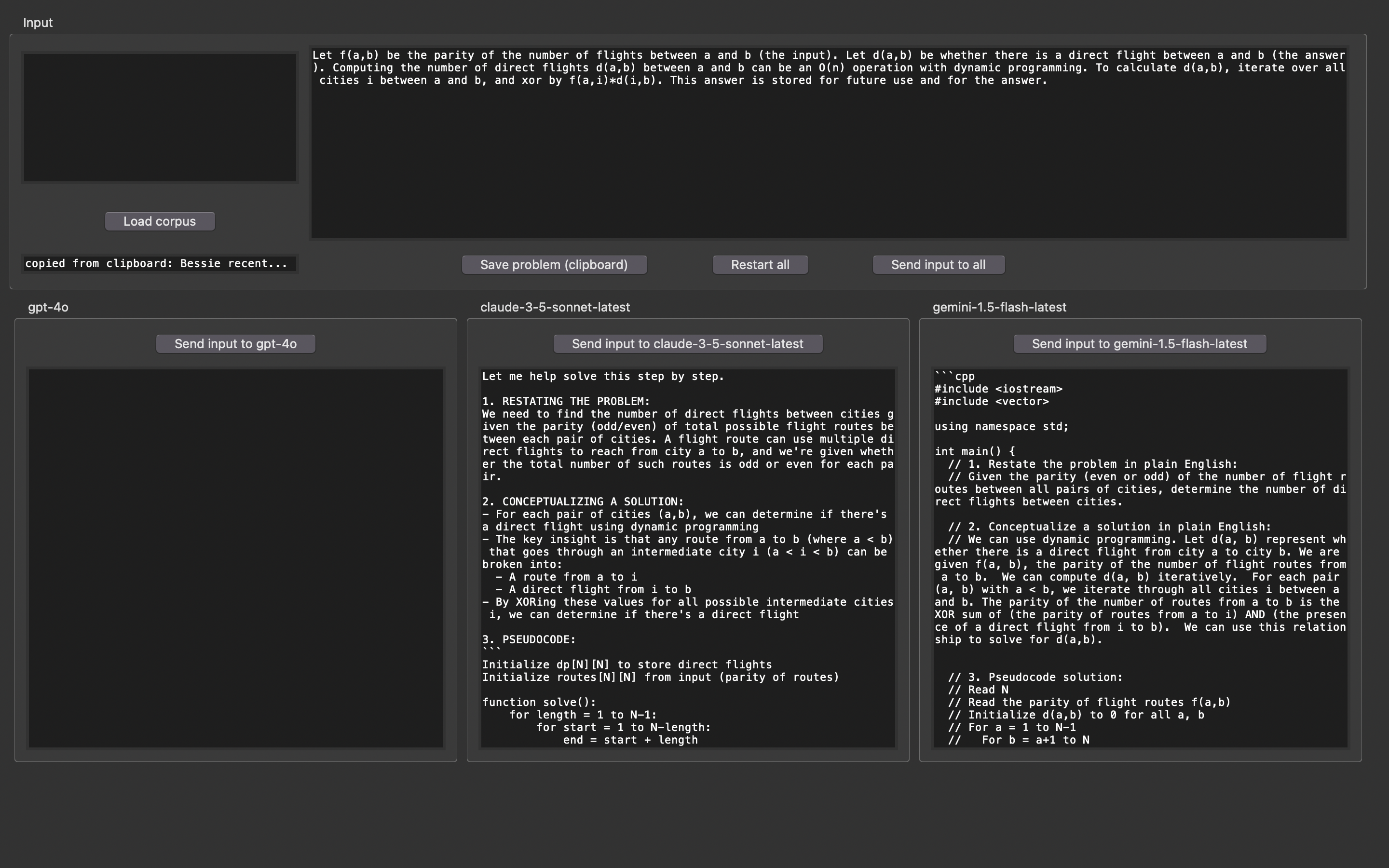} 
    \caption{Outputs asynchronously updating}
\end{figure}

\section{Results}
Having previous experience with a problem gives an advantage when implementing, so six new problems were chosen. USACO's 2023 December Gold Contest and 2024 January Gold Contest were chosen, to try to normalize difficulty where possible.

\begin{table}[h]
    \centering
    \begin{tabular}{|l|c|c|}
        \hline
         & \textbf{Solo (December)} & \textbf{AI assisted (January)} \\  
        \hline
        Problem 1 & 23  & 45  \\
        \hline
        Problem 2 & 66  & 21  \\
        \hline
        Problem 3 & 32  & 26  \\
        \hline
    \end{tabular}
    \caption{Example Table}
    \label{tab:example}
\end{table}

This is an average of a 24\% decrease in implementation time. Note that thinking time, in which the human comes up with an algorithm, is not included, since this merely measures the usefulness of AI in implementing such code. 

Longer solutions clearly correlated with longer implementation times. Comparing between the two contests' model solutions, we see similar lengths in solutions, if not shorter for 2023 December. 
\section{Discussions and Conclusion}

Due to time constraints, we were limited in the amount of data gathered as well as methodology. 
Future work should include

1. Possibly larger text corpuses. Due to the small size of this corpus, human input was feasible. If we include, for example, all accepted solutions to all problems, then it should use a retrieval algorithm. The previous paper used bag-of-words, a somewhat outdated technique. Using encoding models could yield better results.

2. Changes in the prompt. There might be more optimal prompts and possibly different prompts for different models.

3. Investigation of OpenAI's O1 and eventually O3. These, however, approach higher performing levels.

\paragraph{Short-term Implications:}
The capabilities of these models bring forth a discussion about academic integrity in olympiads. The biggest impact is in Olympiad programming, as it is usually done remotely compared to olympiad math or physics. Current models do not have strong enough capabilities on their own to design algorithms at the highest level. For example, in the USA Computing Olympiad, with four divisions of vastly ranging difficulties, models cannot reliably solve problems at any level and zero at the highest level. Most of the tests were also done with problems before these models' training cutoff dates, and in this paper models failed to solve *any* problems past the training cutoff. But as these preliminary results suggest, models offer significant advantage during algorithm implementation. Whether it's allowed as an implementation tool or banned is an ongoing discussion topic.

In recognition of these issues, the current workflow for our AI-assisted approach remains closed-source and is shared only among a small group of individuals. Restricting access to the implementation details helps prevent misuse, at least in the near term, while we explore the broader implications and benefits of AI-driven code generation.

\paragraph{Long-Term Evolution.}
Looking beyond immediate considerations, it is increasingly clear that future generations of programmers will code hand-in-hand with AI assistants. The policies and format of these competitions must adapt to the advances of AI  in order to stay relevant, engaging, and educational. One possibility is the development of new types of problems or scoring systems that focus more on high-level conceptual reasoning and less on rote implementation. Competitions might also emphasize aspects of creativity, human insight, or real-world problem modeling that current AI systems still struggle to automate.

\bibliographystyle{plain}
\bibliography{main}
\end{document}